\documentclass{aa}  

\usepackage{graphicx}
\usepackage{txfonts}
\usepackage{color}

\usepackage{natbib,twoopt}
\usepackage[breaklinks=true]{hyperref} 
\bibpunct{(}{)}{;}{a}{}{,}             
\makeatletter
  \newcommandtwoopt{\citeads}[3][][]{\href{http://adsabs.harvard.edu/abs/#3}%
    {\def\hyper@linkstart##1##2{}%
     \let\hyper@linkend\@empty\citealp[#1][#2]{#3}}}
  \newcommandtwoopt{\citepads}[3][][]{\href{http://adsabs.harvard.edu/abs/#3}%
    {\def\hyper@linkstart##1##2{}%
     \let\hyper@linkend\@empty\citep[#1][#2]{#3}}}
  \newcommandtwoopt{\citetads}[3][][]{\href{http://adsabs.harvard.edu/abs/#3}%
    {\def\hyper@linkstart##1##2{}%
     \let\hyper@linkend\@empty\citet[#1][#2]{#3}}}
  \newcommandtwoopt{\citeyearads}[3][][]%
    {\href{http://adsabs.harvard.edu/abs/#3}
    {\def\hyper@linkstart##1##2{}%
     \let\hyper@linkend\@empty\citeyear[#1][#2]{#3}}}
\makeatother

\begin{document}

   \title{Properties of sunspot light bridges on a geometric height scale}

   \author{S. Esteban Pozuelo\inst{1,2}, A. Asensio Ramos\inst{1,2}, C. J. D\'{i}az Baso\inst{3,4}, 
          \and
          B. Ruiz Cobo\inst{1,2}}

   \institute{Instituto de Astrofísica de Canarias, C/ Vía Láctea, s/n, 38205, La Laguna, Tenerife, Spain\\
              \email{sesteban@iac.es},
              \and
              Departamento de Astrofísica, Universidad de La Laguna, 38206, La Laguna, Tenerife, Spain
         \and
             Institute of Theoretical Astrophysics,
    University of Oslo, %
    P.O. Box 1029 Blindern, N-0315 Oslo, Norway
    \and
    Rosseland Centre for Solar Physics,
    University of Oslo, %
    P.O. Box 1029 Blindern, N-0315 Oslo, Norway}

   \date{Received ; accepted }

\authorrunning{Esteban Pozuelo et al.}

 
  \abstract
 {Investigating light bridges (LBs) helps us comprehend key aspects of sunspots. However, few studies have analyzed the properties of LBs in terms of the geometric height, which is a more realistic perspective given the corrugation of the solar atmosphere.}
   {We aim to shed light on LBs by studying the variation in their physical properties with geometric height.}
   {We used the SICON code to infer the physical quantities in terms of the optical depth and the Wilson depression values of three LBs hosted by a sunspot observed with Hinode/SP in the \ion{Fe}{i} 630~nm pair lines. We also used SIR inversions to cross-check the height variation of the field inclination in the LBs. In both output sets, we performed linear interpolation to convert the physical parameters from optical depth into a geometric height scale in each pixel.}
   {Depending on their general appearance, we classified each LB as filamentary, grainy, or umbral. They appear as ridges that reach different maximum heights, with the umbral LB being the deepest. While the filamentary LB hosts a plasma inflow from the penumbra, the results for the grainy LB are compatible with an injection of hot plasma through convective cells of reduced field strength. Only a few positions reveal hints suggesting a cusp-like magnetic canopy.  
Moreover, strong gradients in the magnetic field strength and inclination usually exhibit enhanced electric currents, with the filamentary LB having remarkably strong currents that appear to be related to chromospheric events.}
   {The height stratification in filamentary and grainy LBs differ, indicating diverse mechanisms at work. Our results are in general incompatible with a magnetic canopy scenario, and further analysis is needed to confirm whether it exists along the entire LB or only at specific locations. Furthermore, this work assesses the usefulness of the SICON code when determining the height stratification of solar structures.}


   \maketitle
%

\section{Introduction} \label{sec:intro}

   Sunspots sometimes host bright and elongated protrusions into the dark umbra, called light bridges (LBs). LBs are related to magnetoconvection in sunspots \citepads{1997ApJ...490..458R, 2004ApJ...604..906R}, similar to umbral dots, bright penumbral grains, or penumbral filaments \citepads{1997A&A...328..682S, 1997A&A...328..689S, 1999A&A...348..621S, 2001A&A...380..714S, 2009A&A...504..575S, 2004ApJ...604..906R, 2006ApJ...646..593R, 2008ApJ...672..684R}. An investigation of these LBs, and of other fine structures, is needed to analyze the relation between mass flows and magnetic fields in sunspots, which is crucial to comprehend the structure of sunspots.
   
Light bridges can arise at different stages of a sunspot's lifetime \citepads[e.g., ][]{1964suns.book.....B, 1970SvA....14...64A, 1987SoPh..112...49G} and are often present in spots with complex magnetic configuration. Furthermore, LBs exhibit different fine structure in the photosphere, a characteristic that is used to classify them \citepads[e.g.,][]{1966AZh....43..480K, 1979SoPh...61..297M, 1993ApJ...415..832S}. For instance, while some LBs appear as extensions of penumbral filaments, most show bright cells resembling quiet-Sun granules. At high spatial resolution, some LBs also reveal a narrow dark lane along the main axis (\citealt{2002A&A...383..275H}, \citealt{2003ApJ...589L.117B}, \citealt{2008ApJ...672..684R}, \citealt{2010ApJ...718L..78R}, and others), which is elevated from the nearby umbra by about 200--450 km \citep{2004SoPh..221...65L}. The appearance of such a dark lane is due to the accumulation of rising plasma at the top of the structure, as in dark cores of penumbral filaments and umbral dots \citepads[e.g., ][]{2006ApJ...641L..73S, 2007ApJ...665L..79B, 2008A&A...488..749R}.

The velocity field in LBs is compatible with the presence of convective motions, where central upward motions are flanked by downflows \citepads[e.g., ][]{1997ApJ...490..458R, 2008ApJ...672..684R, 2008A&A...489..747G, 2010ApJ...718L..78R, 2015ApJ...811..137T}. LBs generally harbor a weaker and more inclined magnetic field compared to the umbra \citepads[][and others]{1969SoPh...10..384B, 1995A&A...302..543R}. Moreover, some studies reveal a cusp-like magnetic canopy over the LB formed by adjacent umbral field lines \citepads[e.g.,][]{1997ApJ...484..900L, 2006A&A...453.1079J, 2016A&A...596A..59F}. Within the canopy, the weak and horizontal magnetic field in the LB becomes stronger and more vertical with height until it is indistinguishable from that in the umbra. Furthermore, field lines inside the canopy may sink at the lateral edges due to the bending and dragging produced by downflowing plasma present there, as \citetads{2014A&A...568A..60L} and \citetads{2016A&A...596A..59F} found only at one edge of the LBs. 

Discontinuities in the geometry of the magnetic field in LBs may cause enhanced electric currents, which could lead to magnetic reconnection events \citepads[e.g.,][]{1997ApJ...484..900L, 2006A&A...453.1079J, 2009ASPC..415..148S, 2011ApJ...738...83S, 2015ApJ...811..137T}. These reconnections are thought to trigger a wide variety of phenomena observed in the chromosphere above LBs (e.g., \citealt{1973SoPh...28...95R}; \citealt{2001ApJ...555L..65A}; \citealt{2003ApJ...589L.117B}; \citealt{2008SoPh..252...43L}; \citealt{2009ASPC..415..148S}; \citealt{2014A&A...567A..96L}; \citealt{2015MNRAS.452L..16B}; \citealt{2016A&A...590A..57R}; \citealt{2021A&A...652L...4L}; \citealt{2021ApJ...907L...4L}). 

The magnetic geometry plays a crucial role in the heating of the chromosphere and transition region (TR), particularly in sunspots, which has garnered significant interest in recent years. For example, \citetads{2021A&A...652L...4L} demonstrated that ohmic energy dissipation by electric currents generated in an LB contributed to an increase in the chromospheric temperature. However, \citetads{2023ApJ...942...62L} analyzed a case where the heating above a grainy LB was too large to be explained by ohmic heating alone, suggesting that additional mechanisms could be at play for the persistent heating at different atmospheric layers above LBs. This study backed the findings of \citetads{2018A&A...609A..73R}, who concluded that LBs are multi-thermal structures that can be observed consistently from the photosphere to the TR. However, these results seem counterintuitive if a cusp-shaped magnetic canopy encloses the LB in the photosphere, as LB-like structures are not expected in the chromosphere and TR. Therefore, we need further investigations on the geometry of the magnetic field in LBs to understand their morphology and impact on the upper atmosphere.
   
In this study we focus on the properties of three LBs in a single sunspot at the photosphere by deriving the physical parameters on a geometric height scale, which has been considered in only a few of the previous investigations \citep[e.g.,][]{2006A&A...453.1079J, 2016A&A...596A..59F}. Given the corrugation of the solar atmosphere, expressing the stratification of the physical parameters on a geometric height scale is essential for a more accurate estimation of the thermal and magnetic properties at different heights, especially when studying structures with steep shapes, such as LBs. Moreover, the transformation to a geometric height scale is also required to estimate the electric current vector, which is an important aspect for understanding LBs and their relation to other events.

\begin{figure}[!t]
\centering
\includegraphics[width=0.5\textwidth, trim={0.2cm 0.2cm 0.2cm 0cm}, clip]{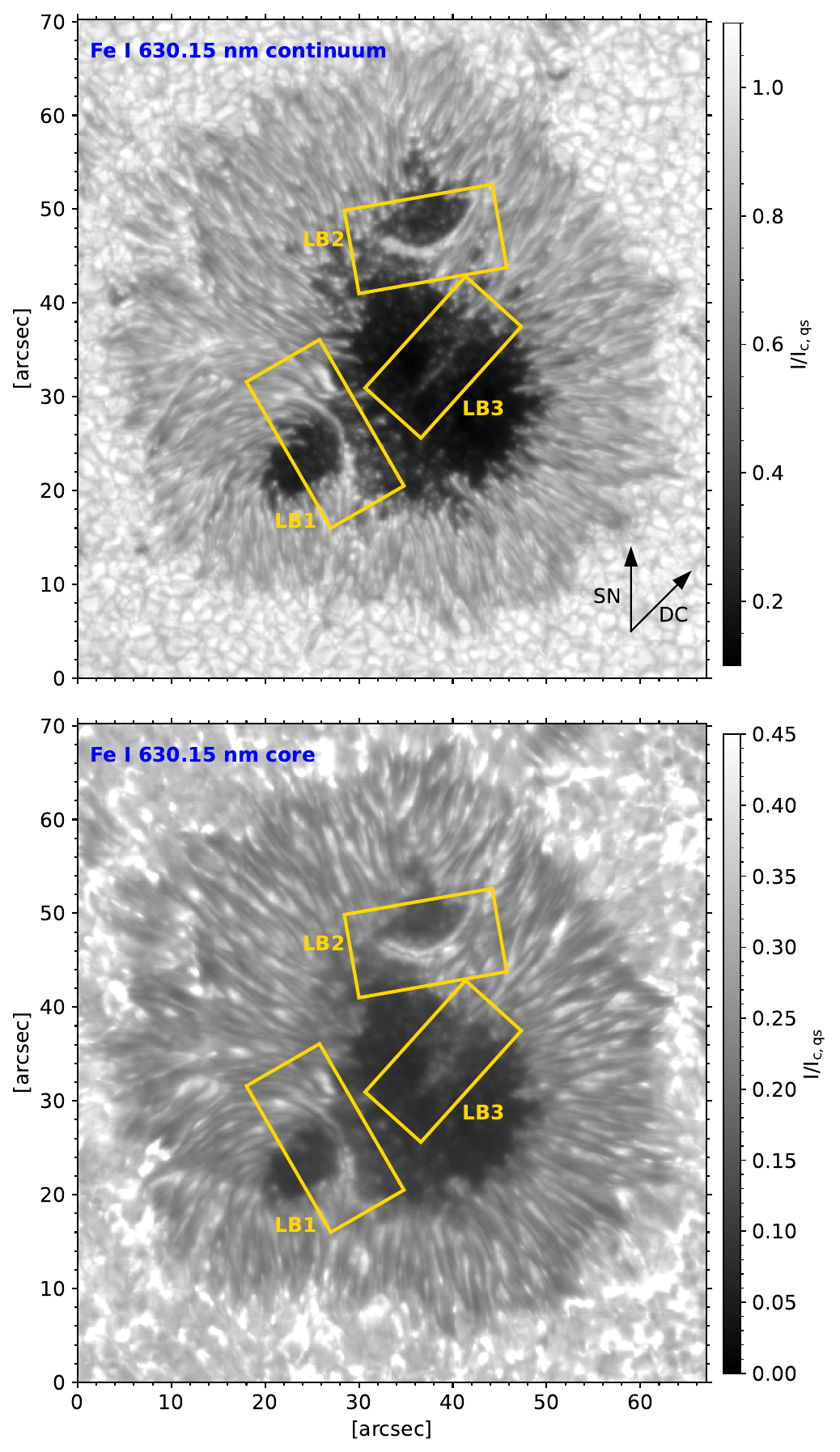}
\caption{Continuum and line core \ion{Fe}{I}~630.15~nm intensity maps (upper and lower panel). The yellow rectangles enclose the analyzed LBs. The arrows in the upper panel point to the disk center (DC) and solar north (SN). $\mathrm{I_{c,qs}}$ refers to the continuum of the averaged quiet-Sun intensity.}
\label{fig:spot}
\end{figure}

    \section{Data} \label{sec:data}

    We analyzed spectropolarimetric data showing the main sunspot of active region (AR) NOAA10953, which is shown in Fig. \ref{fig:spot}. This target was scanned on April 30, 2007, between UT 18:35 and 19:39, with the spectropolarimeter on the Solar Optical Telescope \citep[SOT/SP;][]{2008SoPh..249..167T, 2013SoPh..283..579L} aboard the Hinode satellite \citepads{2007SoPh..243....3K}. Specifically, the spot was located at a heliocentric angle of 13$^{\circ}$ from the disk center during the observation. The data acquisition was performed by sampling the \ion{Fe}{i}~630~nm line pair between 630.089 and 630.327~nm at steps of 21.4~m\AA \, using the normal map mode. The time duration per slit position was 4.8~s.
    The field of view (FOV) is $\sim$164\arcsec$\times$120\arcsec with a spatial sampling of $\sim$0\farcs16 per pixel. The level 1 data used in this work are publicly available at the online archive of the Lockheed Martin Solar and Astrophysics Laboratory.

    \section{Analysis} \label{sec:analysis}

\begin{figure*}
\centering
\includegraphics[width=0.98\textwidth, trim={0cm 0.2cm 1.4cm 1.6cm}, clip]{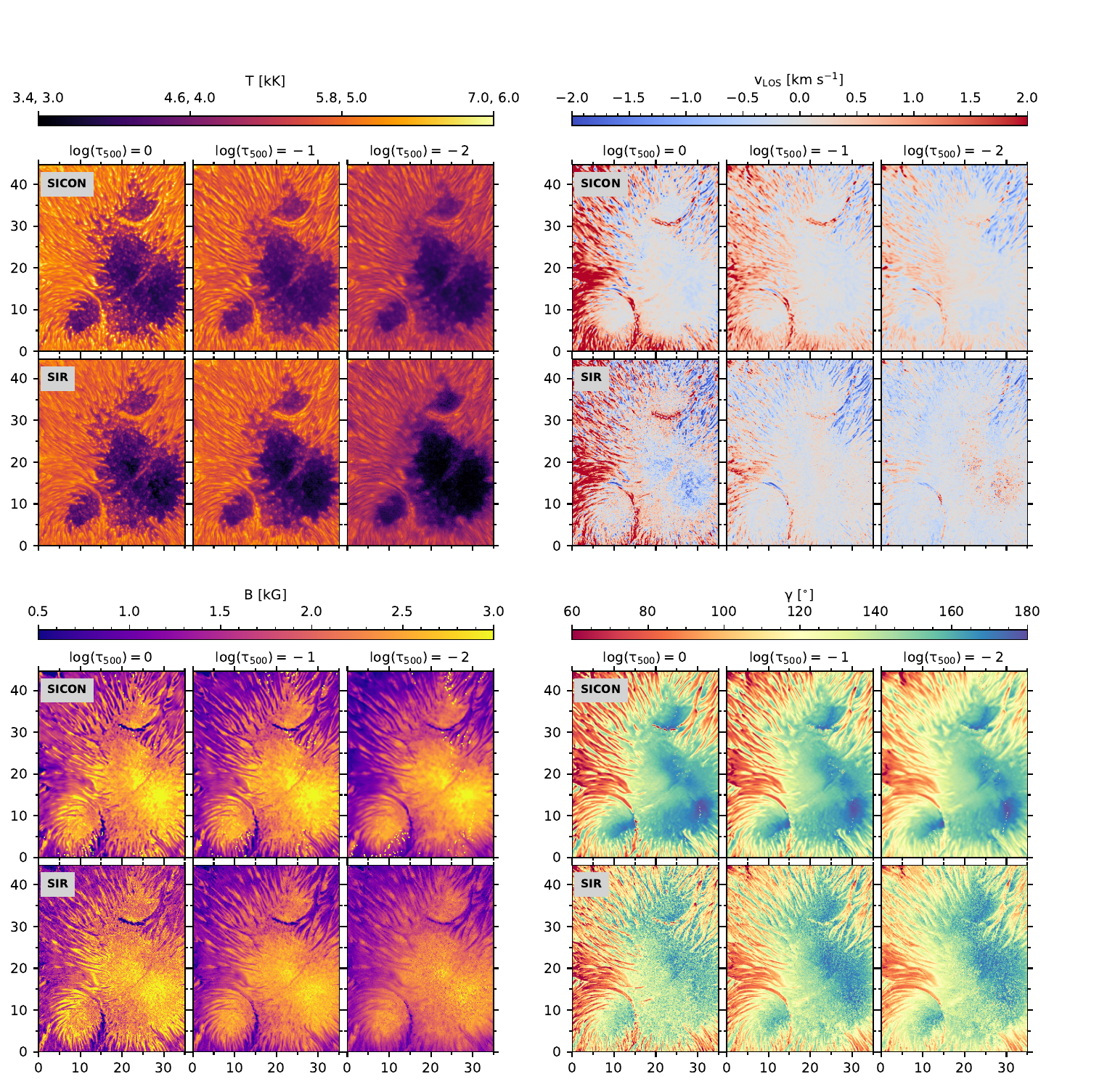}
\caption{Physical quantities retrieved with the SICON and SIR codes at log($\mathrm{\tau_{500}}$)=0, $-$1, and $-$2 in a FOV hosting the analyzed LBs. The top panels show temperature and LOS velocity. The bottom panels show magnetic field strength and inclination. The temperature maps are saturated between 3.4 and 7 kK at log($\mathrm{\tau_{500}}$)=0 and between 3 and 6 kK at log($\mathrm{\tau_{500}}$)=$-$1 and $-$2. The axes are represented in arcsec.}
\label{fig:comp_sicon_sir}
\end{figure*}

\begin{figure*}[!t]
\centering
\includegraphics[width=0.95\textwidth, trim={3cm 0.2cm 0.5cm 2cm}, clip]{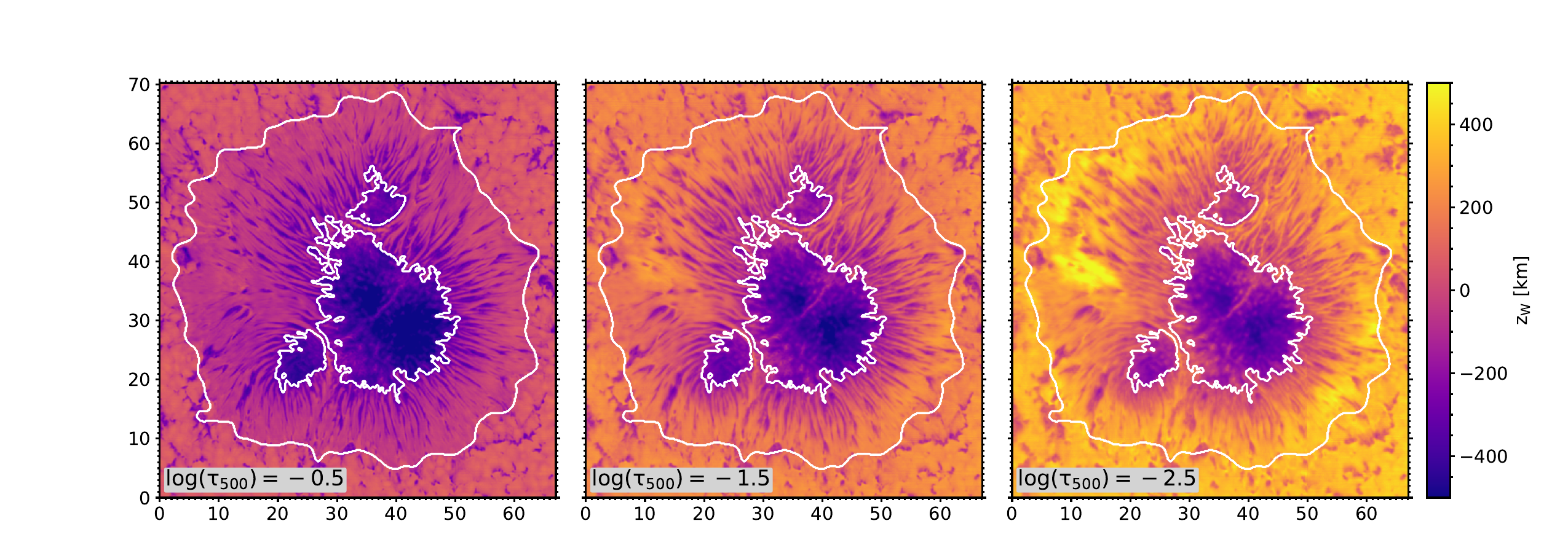}
\caption{Wilson depression maps at $\log (\tau_{500})$=$-$0.5, $-$1.5, and $-$2.5 given by the SICON code. The white contours delimit the umbra and penumbra. The axes are in arcsec.}
\label{fig:zw_map}
\end{figure*}

\begin{figure*}[!t]
\centering
\includegraphics[width=0.94\textwidth, trim={0cm 1cm 2.2cm 1.1cm}, clip]{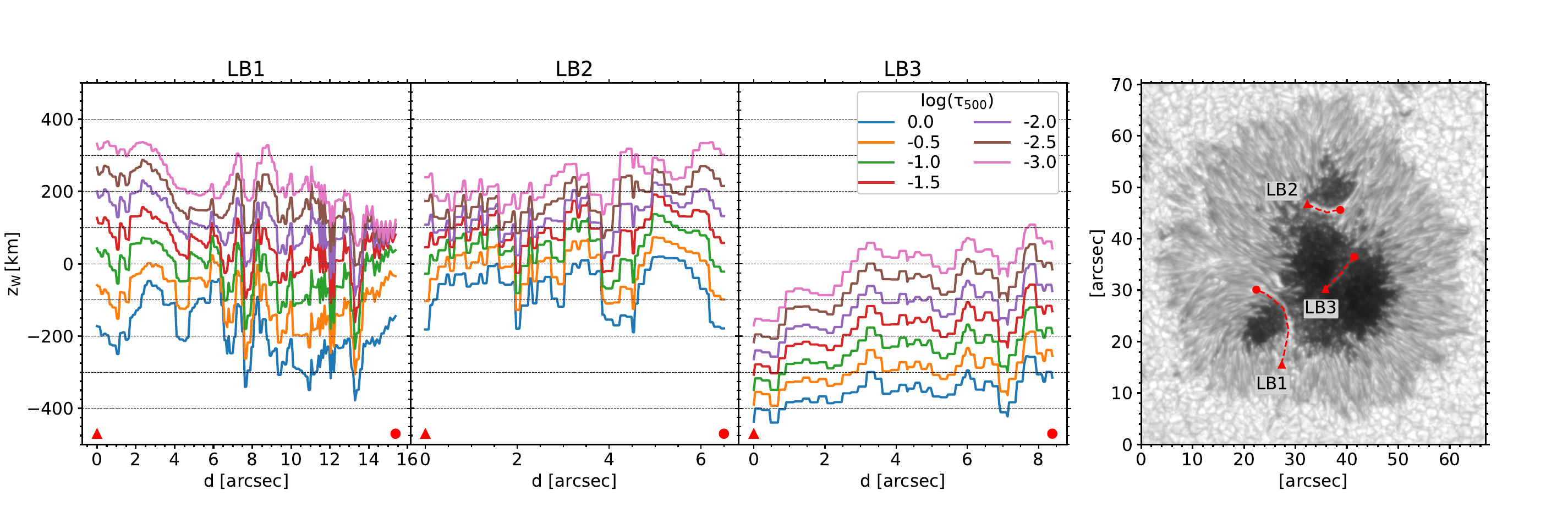}
\caption{Wilson depression at different optical depths along each LB. The \ion{Fe}{I}~630.15~nm continuum intensity map shows the paths used to compute the Wilson depression for each LB and is saturated as the upper panel in Fig.~\ref{fig:spot}. The triangular and circular markers indicate respectively the starting and ending points of the paths. The black dotted horizontal lines (left-hand plots) mark every 100~km.}
\label{fig:zw_all}
\end{figure*}

The rectangles in Fig.~\ref{fig:spot} enclose the three LBs hosted in the observed spot (labeled LB1, LB2, and LB3). LB1 is a filamentary LB, at least most of it, that joins two southern penumbral regions and separates two umbral cores of the same polarity. Its dark lane is only visible in the line core intensity image. LB2 is a grainy LB in the northeast, whose dark lane is more visible in the line core intensity map than in continuum intensity. Finally, LB3 is an umbral LB that appears as an aligned distribution of umbral dots.

We inferred the physical parameters by performing inversions with the \textit{Stokes Inversion based on COnvolutional Neural networks} code\footnote{\url{https://github.com/aasensio/SICON_hinode}} \citep[SICON, ][]{2019A&A...626A.102A}. This code uses a convolutional neural network (CNN) to produce a mapping between the observed Stokes profiles and the physical properties at different layers in the atmosphere. The CNN was trained with two snapshots extracted from three-dimensional (3D) magnetohydrodynamic (MHD) simulations of a sunspot \citepads{2012ApJ...750...62R} and an emerging flux region \citepads{2010ApJ...720..233C} performed with the MURaM code \citepads{2005A&A...429..335V}. The synthetic full-Stokes observations from these training sets were computed using the \textit{Stokes Inversion based on Response functions} code \citep[SIR, ][]{1992ApJ...398..375R}. The synthetic data were processed considering the point spread function (PSF), pixel size, and spectral degradation of Hinode/SP. After extracting randomly a large set of patches from these simulations, the training of the CNN was performed by optimizing a scalar loss function, which measures the difference between each target patch with certain physical conditions and the output of the neural network. Once the training phase is completed, this inversion code can be applied to any Hinode observation. Details on the training sets and the inversion code are described in \citetads{2019A&A...626A.102A}. The code output offers a prompt retrieval of the magnetic and thermodynamic properties in the optical depth scale and an estimation of the Wilson depression \citepads{1774RSPT...64....1W} for each optical depth surface. Additionally, the output parameters are decontaminated from the smearing effect of the Hinode PSF.

We calibrated the LOS velocities using the averaged umbral value as the zero reference. Specifically, we defined the umbra as those pixels with a continuum intensity~$\leq$~0.5~I$\mathrm{_{c,qs}}$, where I$\mathrm{_{c,qs}}$ stands for the continuum of the average quiet-Sun intensity. 

The SICON code gives the output magnetic field vector in the LOS reference frame, which does not coincide with the local reference frame (LRF) as the observations were not performed at the disk center. Assuming a potential field, we solved the 180$^{\circ}$ ambiguity of the magnetic field azimuth following the acute-angle method. Then we transformed the magnetic field to the LRF using the appropriate transformation matrix \citep[e.g.,][]{2018PhDT........98R}. After the disambiguation, we obtained that, in general, the magnetic field in the sunspot is directed radially toward the umbral cores and along the structure in LB1. Finally, because of its relation to the disruption of the magnetic field vector ($\mathbf{B}$), we also inspected the components of the electric current density vector ($\mathbf{J}$), which is computed as

\begin{equation}
    \mathbf{J} \, \mathrm{= \, \frac{1}{\mu_{0}} \left( \nabla \, \times \, \mathbf{B} \right)},
\end{equation}

\noindent where $\mathrm{\mu_{0} \, (=4\pi \times 10^{-7} \, T~m~A^{-1})}$ is the magnetic permeability.

\section{Comparison with results from the SIR code} \label{sec:comp_sir}

Even though neural networks are a powerful tool, results from an inversion code based on this strategy could cast doubts among the community. In this section we compare our results with those inferred by \citetads{2013A&A...549L...4R} with the SIR code. Using a deconvolved version of the same dataset, these authors characterized the atmosphere as a function of the optical depth by considering seven nodes in temperature; five in magnetic field strength, inclination, and LOS velocity; and two in azimuth. We used the same methods to calibrate the LOS velocities and to obtain the magnetic field in the LRF as those applied to the corresponding outputs inferred with SICON (see Sect.~\ref{sec:analysis}).

Figure~\ref{fig:comp_sicon_sir} compares the results given by SICON and SIR at different optical depths. The maps inferred by the former are sharper and show smaller features. LBs generally show similar temperature and magnetic field strength values regardless of the inversion code. Roughly speaking, the LOS velocity and inclination maps given by SICON resemble those inferred with SIR; however, we find important differences in the LBs. Specifically, the LOS velocity maps given by SICON show small but relevant details in LB1 and LB2 that are practically imperceptible when using SIR. In contrast, we can discern the inclination values within LB1 and LB2 in the maps obtained with SIR, while the inclination provided by SICON shows abrupt changes, mainly at log($\mathrm{\tau_{500}}$)=0. Despite these differences, on the whole, the SICON and SIR codes provide comparable scenarios in terms of the optical depth. In the following we describe the results obtained with the SICON code in terms of the geometric height. Furthermore, we also analyze independently the values inferred for the inclination at different heights with SIR.

\section{Atmospheric stratification in geometric height} \label{sec:resgeo}

The mapping between geometric height and optical depth for a given pixel depends on the specific physical
properties at that location. Consequently, the surface at a fixed constant optical depth is corrugated.
The inference of this mapping is challenging when only using spectropolarimetric methods, and
previous approaches have relied on using additional constraints. \cite{2010ApJ...720.1417P} were arguably the first to estimate the mapping by minimizing the divergence of the magnetic field vector and the deviations from static equilibrium. This approach was later considered by \cite{2020A&A...635A.202L}. 

Some inversion codes directly provide the stratification of the atmospheric parameters on a geometric height scale. For instance, the \textit{MHD-Assisted Stokes Inversion} method \citep[MASI, ][]{2017ApJS..229...16R} uses degraded MHD simulations to assign in each observed pixel the atmosphere model related to the synthetic Stokes profiles that best fit the observed ones. The resulting mosaic is the initial condition of new iterative MHD simulations to assure physical consistency throughout the FOV. Alternatively, \citetads{2019A&A...629A..24P} presented the FIRTEZ-dz code, which directly solves the radiative transfer equation for polarized light on a geometric height scale.

\begin{figure*}[!h]
\centering
\includegraphics[height=0.87\textheight, trim={0cm 0.4cm 0.4cm 5.2cm}, clip]{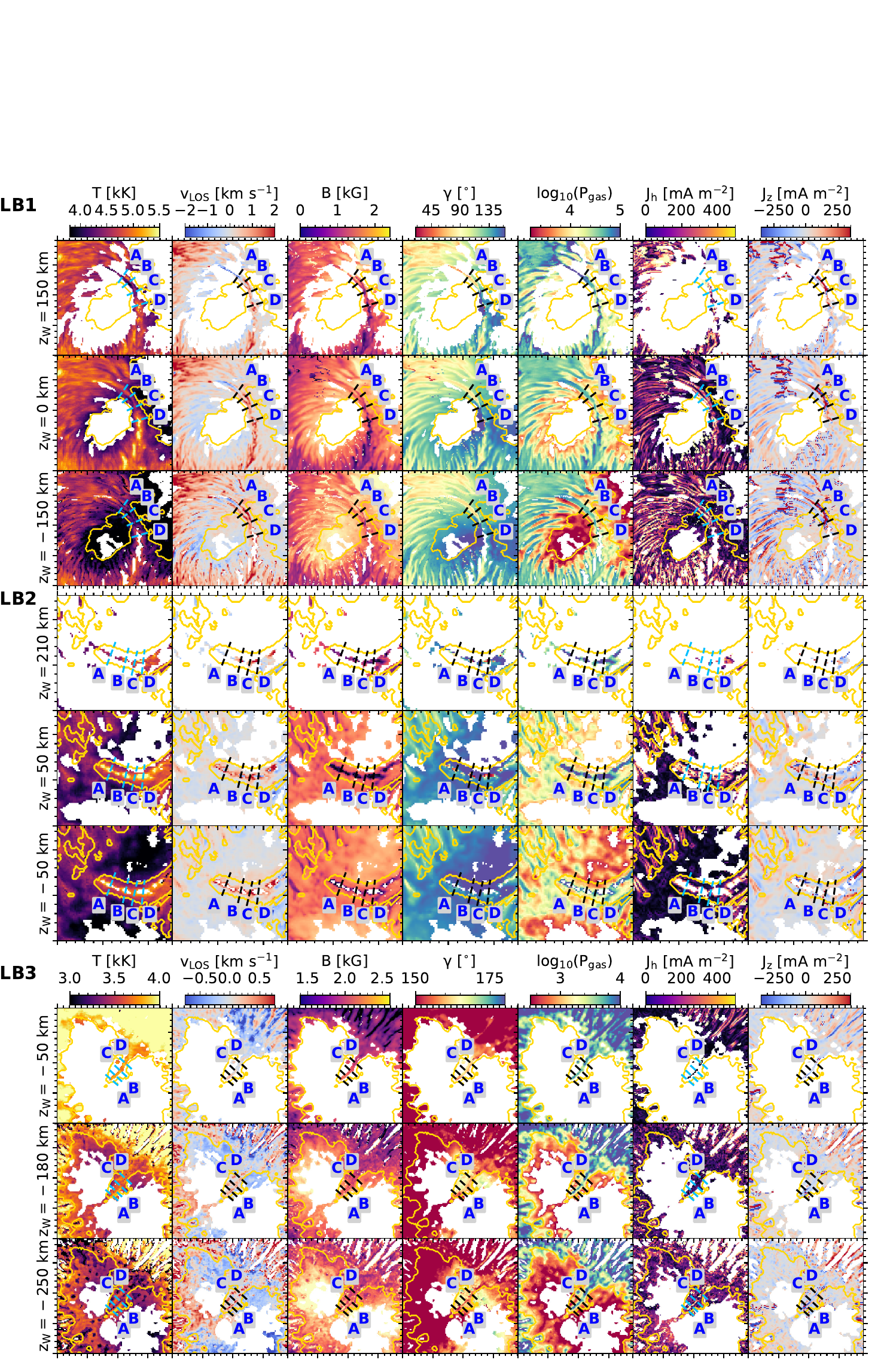}
\caption{Physical quantities inferred at different heights for each LB. From left to right: temperature, LOS velocity, magnetic field strength, inclination, logarithm of the gas pressure, horizontal and vertical components of the \textbf{J} vector. For visualization purposes, the yellow contours delimit umbral pixels with \ion{Fe}{I} 630.1~nm line core intensity $<$ 0.25~I$\mathrm{_{c,qs}}$ (LB1 and LB2 panels) and $<$ 0.12~I$\mathrm{_{c,qs}}$ (LB3 panels). Slits A--D mark the position of the vertical cuts shown in Figs.~\ref{fig:param_vertcut_lb1}, \ref{fig:param_vertcut_lb2}, and \ref{fig:param_vertcut_lb3}. Each major tickmark represents 5\arcsec. No information was retrieved within the blank areas.}
\label{fig:param_z_lbs_mono}
\end{figure*}

In the case of the SICON code, it learns this mapping from the simulations used as training sets. Theoretically, one should train the CNN with synthesis performed at the same heliocentric distance as the observations to assure consistency between the two datasets. However, \citetads{2019A&A...626A.102A} showed the seamless performance of the SICON code after comparing the outputs obtained from ARs located at about 14$^{\circ}$ and 43$^{\circ}$ from the disk center, respectively, to previous results in the literature. We thus expect little deviation from the results we would infer using training sets computed with the same heliocentric angle as present in the observations (13$^{\circ}$).

Figure~\ref{fig:zw_map} displays the output Wilson depression values for $\log(\tau_{500})$=$-$0.5, $-$1.5, and $-$2.5. The height $\mathrm{z_{W}}$=0 refers to the average height of the quiet Sun at $\log(\tau_{500})$=0. We infer an expected height increase as the optical depth decreases. Our geometric heights in the umbra at a $\log(\tau_{500})$ of $-$0.5 and $-$1.5 resemble those obtained by \citetads{2020A&A...635A.202L} and are also consistent with those calculated by \citetads{2021A&A...647A.190B}. We also observe a conspicuous patch showing heights $>$ 400~km at $\log(\tau_{500})$=$-$2.5 (see coordinates (X, Y)=(35\arcsec, 55\arcsec)), which may be related to enhanced activity above the AR \citep[see, e.g.,][]{2010ApJ...715.1566C}. 

All maps in Fig.~\ref{fig:zw_map} show the analyzed LBs. LB1 covers heights from $-$150~km to 200~km between $\log(\tau_{500})$=$-$0.5 and $-$2.5. At the same optical depths, the location of LB2 and LB3 ranges between $-$100 and 200~km and from $-$380 to 0~km, respectively. In addition, Fig.~\ref{fig:zw_all} displays how the inferred geometric heights vary along the LBs at different optical depths. Specifically, we represent the heights retrieved along the dark lane of each LB as seen in the \ion{Fe}{I}~630.15~nm line core intensity map. The geometric heights in LB1 and LB2 are similar, though we can see deeper at some positions of LB1. Optical depths above 0.1 correspond to heights below 0~km (reaching down to $-$350~km), whereas above $\log(\tau_{500})$=$-$1.5 heights are above 0~km (up to 150~km). In contrast, geometric heights in LB3 are below 0~km, except for $\log(\tau_{500})$=$-$3. On average, LB3 is located 200~km deeper than the others.

\begin{figure}[!t]
\centering
\includegraphics[width=0.49\textwidth, trim={0.4cm 0.5cm 0.1cm 0.5cm}, clip]{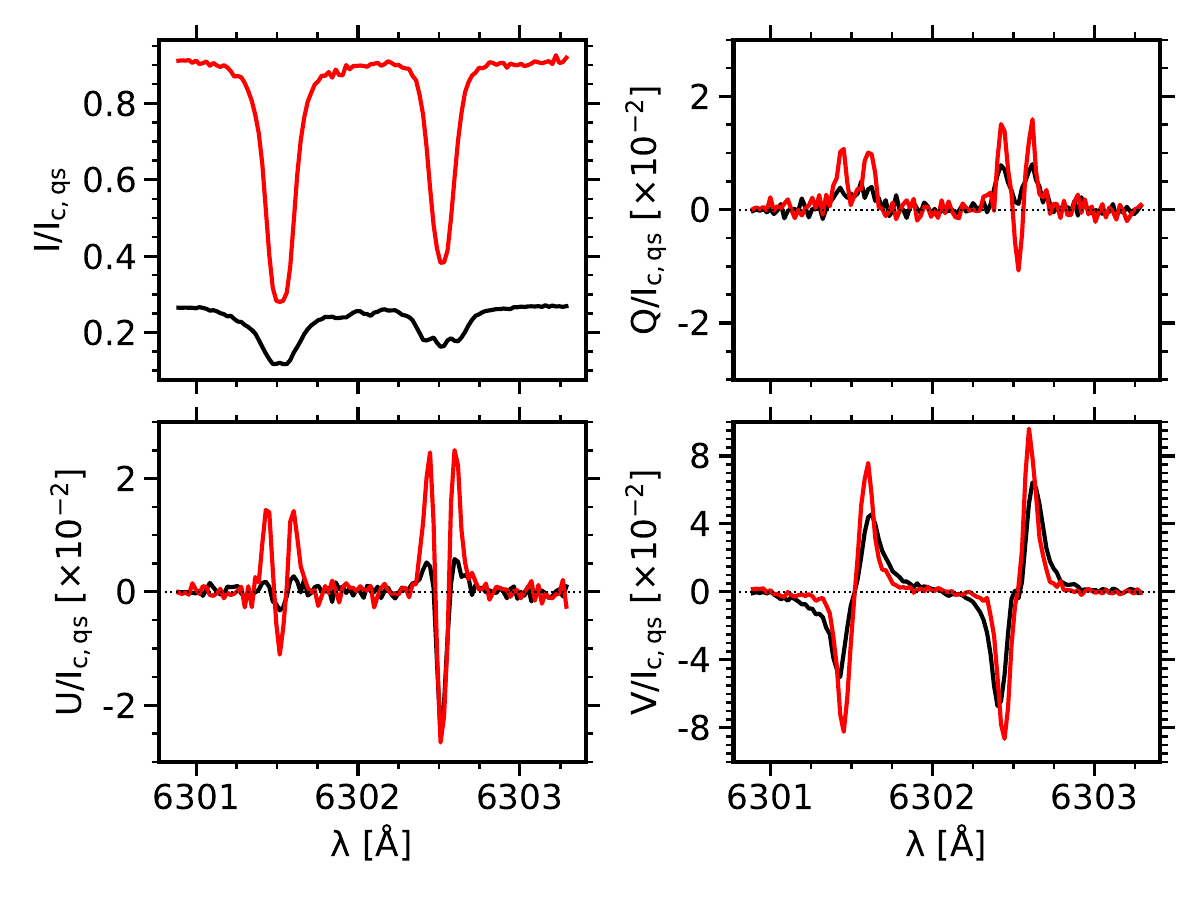}
\caption{Stokes profiles emerging at the center of cut D in LB1 (red lines), where the retrieved inclination indicates a field reversal while the associated profiles do not show a change in the sign of the lobes. For comparison, the Stokes profiles observed at an umbral position are also plotted (black lines). The horizontal dotted lines indicate zero polarization signals.}
\label{fig:stokes_pix_lb1}
\end{figure}

The corrugated mapping between optical depth and geometric height can be inverted so that
we end up with the physical parameters in a geometric cartesian coordinate system. We do so by performing linear interpolation in the allowed range of heights for each pixel. 
In the following we present the variation in the physical quantities inferred for each LB with the geometric height.

Figure~\ref{fig:param_z_lbs_mono} displays the physical quantities inferred for each LB at three geometric heights. In particular, we selected these geometric heights to show the lower, middle, and upper regions of the height stratification retrieved for each LB. A yellow contour delimits the area hosting the LBs for ease of visualization. 

First, we describe the results obtained for LB1 (top panels). On first inspection, we observe that the information retrieved at $\mathrm{z_{W}}$=$-$150~km mainly comes from the northern part of the LB, while we need to examine higher up to see the rest of the LB. This height difference suggests a slope along LB1, as also evident from the left panel of Fig.~\ref{fig:zw_all}.

Furthermore, the physical parameters along LB1 vary significantly. In the northern part, a cold lane between two hotter lateral paths is visible in the temperature maps until $\mathrm{z_{W}}$=150~km, when only the cold lane is discernible. This part of LB1 shows a central blueshift of $-$1.2~km~s$\mathrm{^{-1}}$ and lateral redshifts of 1.5~km~s$\mathrm{^{-1}}$ at --150~km height, while we only infer a central blueshift of $\sim-$1~km~s$\mathrm{^{-1}}$ higher up. Although the temperature and LOS velocity patterns in this part of LB1 resemble those in a penumbral filament in the center-side penumbra, this flow channel is located in the limb-side penumbra suggesting an inflow of plasma from the penumbra protruding into the umbra (as suggested by \citealt{2011ApJ...738...83S} in the same LB), which was falling at the middle of the structure at the moment of the observations. On the other hand, the temperature map at $\mathrm{z_{W}}$=0~km shows patches in the southern part of LB1 that are 800~K hotter than their surroundings. In the corresponding dopplergram, we mainly observe redshifted LOS velocities and some blueshifted (or at rest) locations that smooth with increasing height. In contrast to the northern part of LB1, these signatures in temperature and LOS velocity resemble those in the quiet-Sun granulation. Furthermore, we do not observe a strict correlation between the patterns detected in temperature and LOS velocity in either part of LB1. 

Regarding the magnetic field, the northern part of LB1 shows horizontal (or slightly reverse) fields which are strong (1--2~kG). In contrast, fields are much weaker inside the hot patches found in the southern part (<500~G). The height variation of the field strength also differs along LB1. Whereas the field strength decreases with height in the northern part, it increases by 200--400~G in the southern part 150~km above. Meanwhile, in the northern part, the field at the center of the LB has opposite polarity to that of the sunspot and tends to be more vertical with height, which appears as a decrease in  the field inclination with increasing height. Unfortunately, the variation in the field inclination in the southern part of the LB is unclear. There, we infer changes of polarity that seem unreal according to the emerging Stokes $V$ profiles (red line in Fig.~\ref{fig:stokes_pix_lb1}).

The differences in the magnetic field along LB1 influence the electric current density values. We computed the horizontal electric current density ($J_{h}$) using the $x$ and $y$ components of the electric current density vector as $J_{h} = \sqrt{J_{x}^2+J_{y}^2}$, while the vertical electric current density ($J_{z}$) simply stands for the $z$ component. First, we focus on the currents found in the northern part of LB1. The $J_h$ values are, in general, similar to those in the penumbra. At $z_{W}$ = $-$150~km, we observe two elongated paths of strong $J_h$ ($\sim$ 400--650~mA~$\mathrm{m^{-2}}$) that partially overlap the LB. Moreover, the northern part of LB1 and the penumbra show a similar distribution of $J_z$ formed by two elongated paths with positive and negative currents on either lateral side. We infer absolute values of $J_z$ ranging between 100--350~mA~$\mathrm{m^{-2}}$, which again overlap partially with the northern part of LB1. The $J_h$ and $J_z$ components both tend to decrease with increasing height; this trend is particularly abrupt for the positive $J_z$ values.  

The southern part of LB1 also shows strong $J_h$ and $Jz$ components. At $z_W$ = 0, we observe an elongated shell with $J_h$ values of 400--600~mA~$\mathrm{m^{-2}}$ that partly overlaps the hot patches found in the temperature map. However, these two parameters may not be correlated as nearby penumbral locations show similar $J_h$ values that are not associated with a temperature increase. Moreover, the southern part of LB1 does not show the same distribution of $J_z$ as the northern part. Instead, we infer rather positive $J_z$ values of $\sim$200~mA~$\mathrm{m^{-2}}$, which again partially coincide with the hot patches. Both $J_h$ and $J_z$ usually decrease with height, though it is not easy to discern as the region diminishes severely with height. 

The middle panels of Fig.~\ref{fig:param_z_lbs_mono} portray the height variation of physical quantities for LB2, which has a uniform behavior along the LB compared to LB1. At $\mathrm{z_W}$=$-$50~km, we mainly infer information from the lateral edges of LB2, where redshifts of $\sim$1~km~s$^{-1}$ coincide with hotter locations compared with the umbral surroundings. At 100~km above, the center of the LB exhibits compact hot patches with central blueshifts and higher gas pressure. Higher up, we only detect a lane at the top of LB2 with lower temperature, higher gas pressure, and mostly blueshifted LOS velocities. Nonetheless, the height stratification inferred for LB2 is not smooth, as Sect.~\ref{sub:complete_lb2} shows.

\begin{figure}[!t]
\centering
\includegraphics[width=0.49\textwidth, trim={0.4cm 0.5cm 0.1cm 0.5cm}, clip]{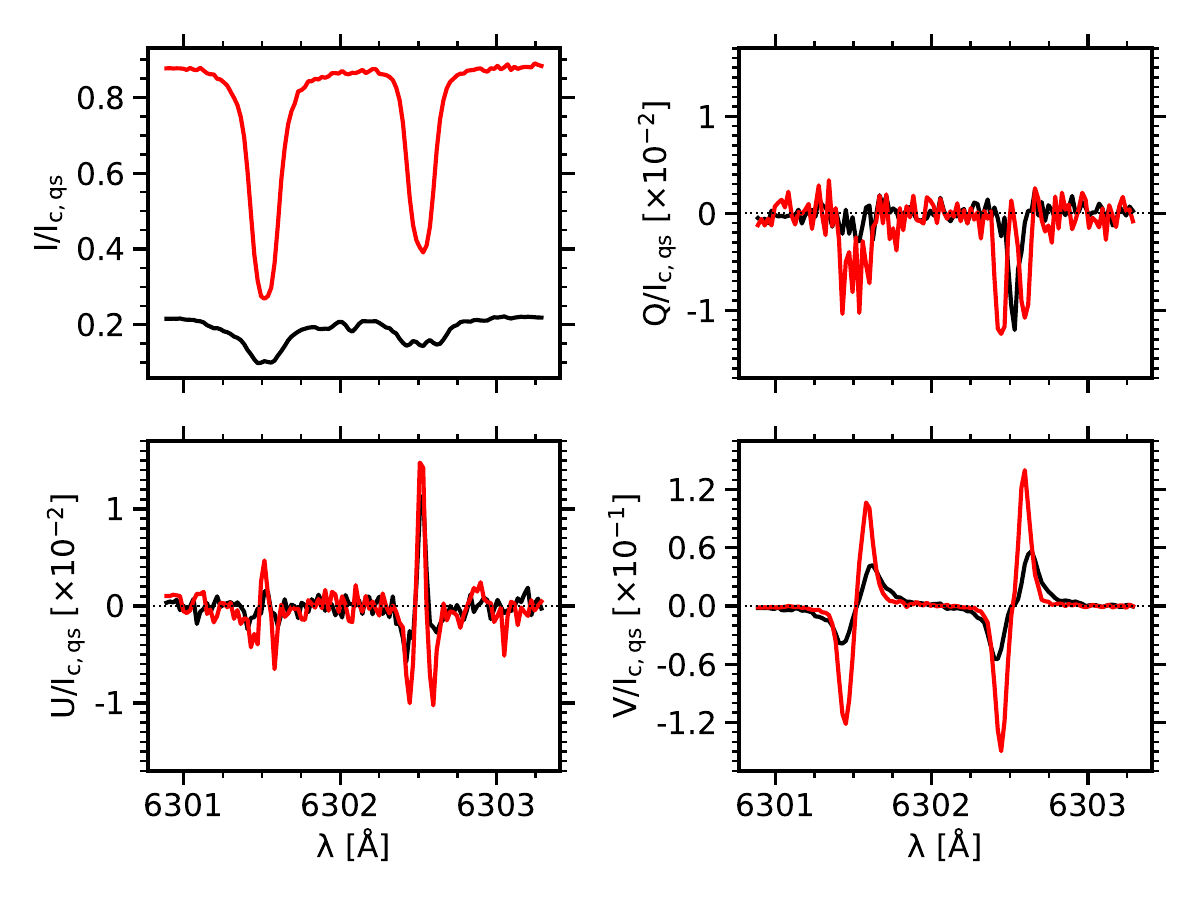}
\caption{Stokes profiles emerging at the center of cut A in LB2 (red lines), where we infer a field reversal according to the retrieved inclination, while the observed Stokes profiles do not show a change in the sign of the lobes. For comparison, the Stokes profiles observed at an umbral position are also plotted (black lines). The horizontal dotted lines indicate zero polarization signals.}
\label{fig:stokes_pix_lb2}
\end{figure}

The magnetic field in LB2 is arranged in compact patches of reduced field. Similar to LB1, some positions show reverse fields that may be unreal, as revealed by the Stokes $V$ profile emerging from an affected pixel (red line in Fig.~\ref{fig:stokes_pix_lb2}). Considering those positions not showing field reversals, patches with reduced field are surrounded by strong $J_h$ values of 400--600~mA~m$\mathrm{^{-2}}$ at $z_{W}$ = 50~km. Meanwhile, the $J_z$ currents are of $\pm$100--200~mA~m$\mathrm{^{-2}}$, with the positive values being predominant. We cannot discern the height variation of $J_h$ and $J_z$.

Interestingly, we find some differences when comparing the results for LB2 found in optical depth and geometric height scales. In optical depth (see Fig.~\ref{fig:comp_sicon_sir}), LB2 appears as a compact and hot structure that faintly shows a central colder lane at log($\tau_{500}$)=0, which is more clear at greater optical depths. In addition, the LOS velocity pattern is compatible with the presence of convective motions at the shown optical depths, although LOS velocities are smoother at log($\tau_{500}$)$>$0. Meanwhile, LB2 shows a reduced and more horizontal magnetic field than the umbra at log($\tau_{500}$)=0, which is stronger and more vertical at greater optical depths. However, when using a geometric height scale (Fig.~\ref{fig:param_z_lbs_mono}), the grainy character of this LB stands out, so its patches look sharper. In addition, the variation in the physical parameters appears dissected along the height stratification, particularly in the temperature and LOS velocity maps.

Finally, the bottom panels of Fig.~\ref{fig:param_z_lbs_mono} show the physical parameters for LB3. Since we cannot identify apparent features closer to the penumbra, we focus on the outer end of LB3 (the furthest from the penumbra). Similar to LB2, we observe differences in the results obtained in optical depth and geometric height scales, mainly in temperature. While LB3 appears as a chain of aligned bright grains in the temperature maps at $\log(\tau_{500})$=0 and $-$1 (see Fig.~\ref{fig:comp_sicon_sir}), it hosts a cold lane along its main axis at $\mathrm{z_W}$=$-$250~km. Temperatures tend to smooth out with increasing height. At the same time, the information from its borders vanishes until we only observe a region above the cold lane with greater temperature. In particular, we find that the cold lane coincides with higher gas pressure and redshifted LOS velocities ($\sim$0.6~km~s$\mathrm{^{-1}}$) at $-$250~km height. 

Furthermore, at $\mathrm{z_W}$=$-$250~km, LB3 shows a lane with strong fields (about 2.5~kG) surrounded by weaker fields ($\sim$0.5~kG less) that coincides with the variation in temperature. Higher up, we infer smoother field strengths from an ever reduced area. Moreover, the field inclination differs significantly on either side of the LB, so fields are more horizontal on the side showing the cold lane in temperature. Regarding $J_h$ and $J_z$, at $z_W$ = $-$250~km, we infer strong $J_h$ values of 300--450~mA~m$\mathrm{^{-2}}$ on a lateral side with prominent $J_z$ values of $\sim$100~mA~m$\mathrm{^{-2}}$. As height increases, $J_z$ values smooth out while $J_h$ decreases.

\section{Height stratification along the LBs}
\label{sec:complete_stratification}

The previous section shows the results for each LB at the lower, middle, and upper regions of their stratifications. However, some aspects are better seen by inspecting all the height ranges. This section analyzes the complete stratification at particular positions describing each LB.

\subsection{Light bridge 1}
\label{sub:complete_lb1}

\begin{figure*}[!t]
\centering
\includegraphics[width=0.98\textwidth, trim={0.1cm 0.2cm 2.3cm 0.5cm}, clip]{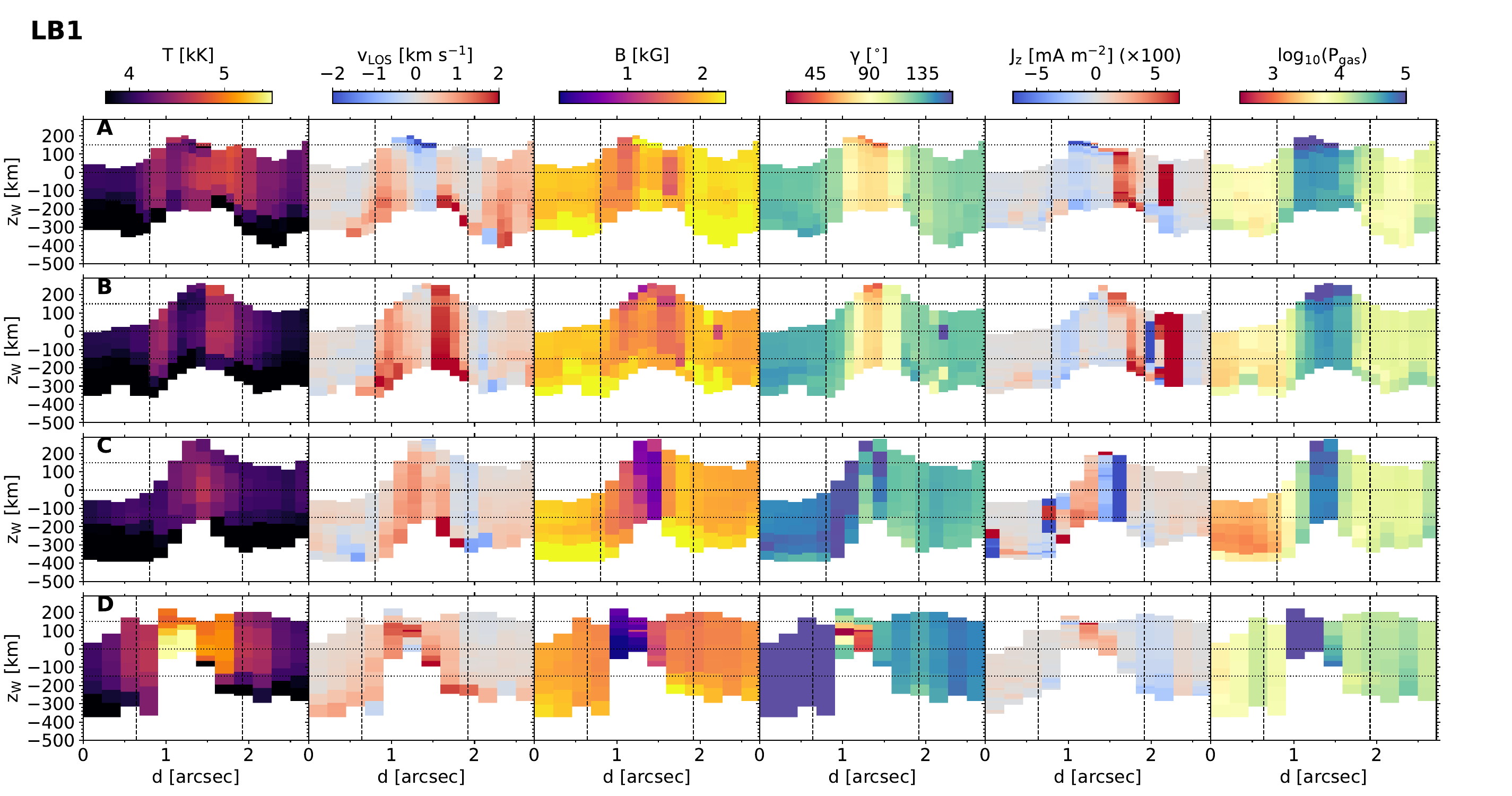}
\caption{Variation in physical quantities with geometric height along the cuts A--D in LB1. From left to right: temperature, LOS velocity, magnetic field strength, inclination, vertical current density, and logarithm of the gas pressure. The horizontal lines trace the heights shown in the upper panels of Fig.~\ref{fig:param_z_lbs_mono}. The vertical dashed lines help distinguish pixels inside the LB.}
\label{fig:param_vertcut_lb1}
\end{figure*}

\begin{figure*}[!t]
\centering
\includegraphics[width=0.98\textwidth, trim={0.1cm 0.2cm 2.3cm 0.5cm}, clip]{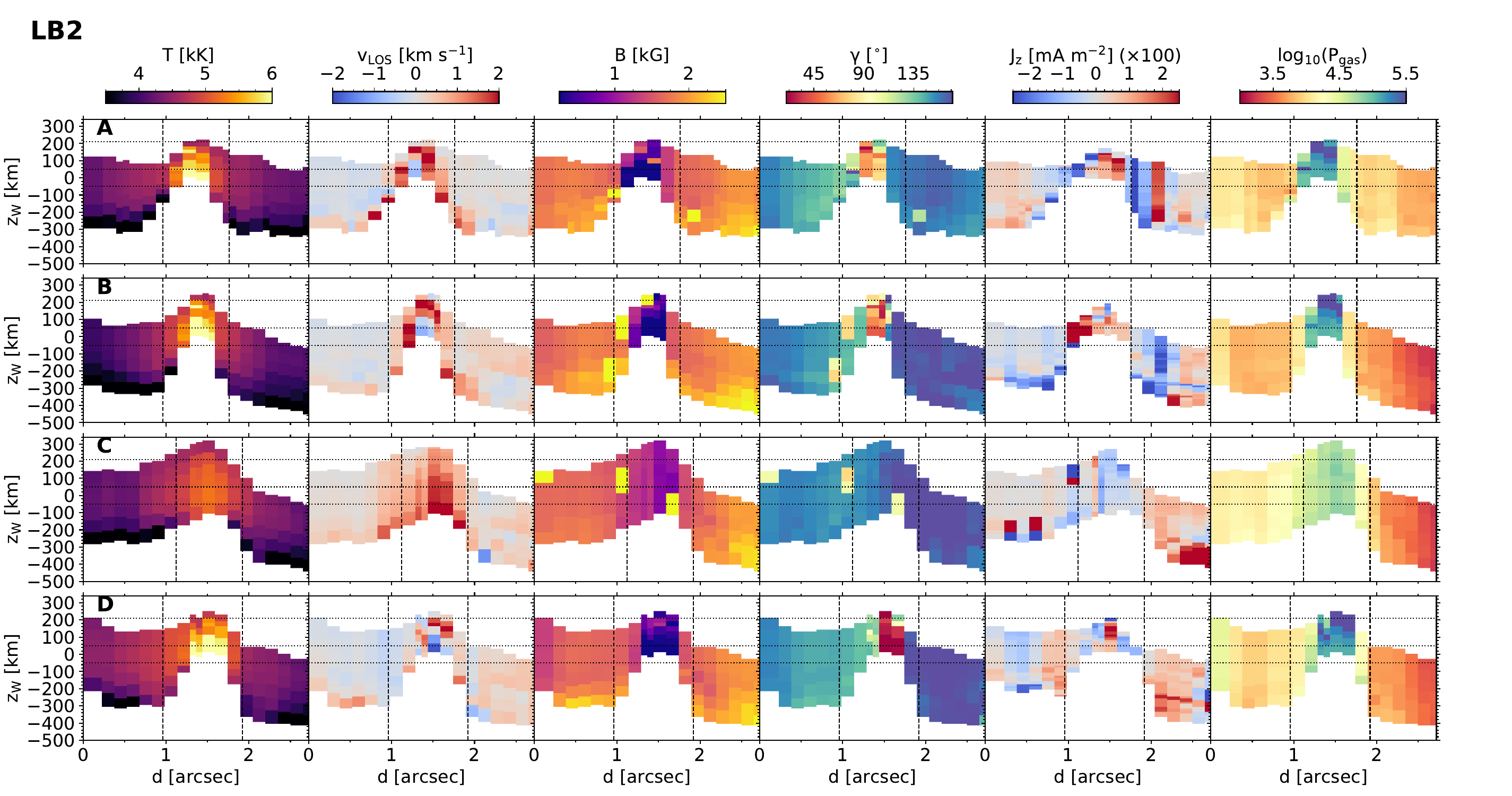}
\caption{Same as Fig.~\ref{fig:param_vertcut_lb1}, but for the cuts A--D in LB2. The horizontal lines indicate the heights shown in the middle panels of Fig.~\ref{fig:param_z_lbs_mono}.}
\label{fig:param_vertcut_lb2}
\end{figure*}

Figure~\ref{fig:param_vertcut_lb1} shows the height variation of multiple parameters along the four slits marked A, B, C, and D in the upper panels of Fig.~\ref{fig:param_z_lbs_mono}. Although all positions show a cuspidal shape, we identify differences among them. In the northern part (cuts A and B), the center of LB1 shows more blueshifted LOS velocities than at the lateral edges, primarily at cut A. Moreover, the central region coincides partially with a colder lane, higher gas pressure, and weaker and horizontal (or slightly reverse) magnetic fields. As height increases, we find more blueshifted LOS velocities, lower temperature, and higher gas pressure. Although the magnetic fields tend to weaken with increasing height, they are usually above 500~G. These variations are not abrupt except at the top of the LB. On the other hand, the field inclination does not show noticeable changes with height except for field reversals at the top of the LB. Considering its position in the sunspot and the results on temperature, LOS velocity, gas pressure, and field strength, the imprints found in the northern part of LB1 may indicate hot plasma flowing horizontally and sinking at the lateral edges. Regarding $J_{z}$, its characteristic distribution persists with height, with the positive values dominating the structure.

Between the northern and southern parts of the LB (cut C), we find a temperature rise that partially coincides with redshifted LOS velocities, reduced magnetic field, and higher gas pressure. Usually, these parameters smoothen with increasing height. These imprints may suggest plasma sinking at this position after flowing along the northern part of LB1. Furthermore, we find a transition to more vertical fields with height (with $\gamma$ > 135$^{\circ}$), which may agree with the cusp-like canopy scenario. However, these vertical fields are not as strong as expected in such a case. Furthermore, we find a reverse distribution of $J_z$ compared to the northern part (cuts A and B).

In the southern part (cut D), the center of the LB shows higher temperature and gas pressure while the magnetic field appears dramatically reduced. These variations partially coincide with a deep and small blueshift. This particular location thus may be consistent with the presence of convective motions. Unfortunately, the reversals inferred at cut D may be unreal, so we cannot extract any clear conclusion from the $J_{z}$ values. As height increases, we can only distinguish a decrease in temperature and more redshifted LOS velocities.

To summarize, the height stratification of the northern and southern parts of LB1 differ conspicuously. The former not only shows smoother gradients but also appears seemingly wider. This can be seen in Fig.~\ref{fig:param_vertcut_lb1}, where the floor of the northern and southern parts (located at $-$250 and $-$50~km, respectively) show a gradient of 200~km, while that of the nearby umbra (at $-$350~km) barely changes. Furthermore, according to the temporal evolution of this LB \citepads[shown in][]{2011ApJ...738...83S}, differences along LB1 may be related to changes in the plasma inflow of the northern part (see also Figs. 4 and 6 of \citealt{2008SoPh..252...43L}). The southern part of LB1 might be due to injections of plasma from below through remaining conduits with locally weaker fields. Specifically, we plan to analyze the height stratification and temporal evolution of this LB in a coming study.  

\subsection{Light bridge 2}
\label{sub:complete_lb2}

Light bridge 2 reveals its grainy morphology when we analyze it in terms of the geometric height. As we did for LB1, we selected four cuts along LB2 to inspect the height stratification inside and outside the hot patches, which is shown in Fig.~\ref{fig:param_vertcut_lb2}.

Positions coinciding with a temperature rise (cuts A, B, and D) show deep blueshifted LOS velocities accompanied by lateral redshifts, as well as higher gas pressure and strongly reduced fields compared to their surroundings. These signatures seem compatible with the presence of convective flows. Regarding the height variation, we observe a decrease in temperature and an increase in gas pressure, but it is difficult to discern how the LOS velocity varies with height. The reason for this is the appearance of sudden changes at $z_W$ = 120--200~km, possibly due to an undesirable effect of the conversion from the optical depth scale to the geometric height. In this process, the range of optical depths in LBs is compressed into stratifications of $\sim$250~km height at some positions, and consequently, a parameter having a smooth stratification in optical depth can show sudden changes in geometric height. Nonetheless, we infer small and weak blueshifts at 200~km height. Moreover, the magnetic field becomes stronger at some positions (see cut A), while it barely changes at others. The inclination values show field reversals that may be unreal (see Sect.~\ref{sec:resgeo}), so we cannot conclude anything about this parameter. Positive currents are predominant in the stratification of $J_z$, though the greatest values in cuts A and B are on opposite sides. 

On the other hand, cut C is between two hot patches. There, the LB also shows a rise in temperature and gas pressure and a decrease in field strength with respect to the nearby umbra. However, these changes are less prominent than those in cuts A, B, and D, where we retrieve temperatures around 1~kK higher and field strengths half as strong as that at cut C. In addition, in contrast to the other cuts, the LB exhibits rather redshifted LOS velocities coinciding with pixels with higher temperature. Moreover, negative $J_z$ currents are dominant at this position.   

\begin{figure*}[!t]
\centering
\includegraphics[width=0.98\textwidth, trim={0.1cm 0.1cm 2.3cm 0.5cm}, clip]{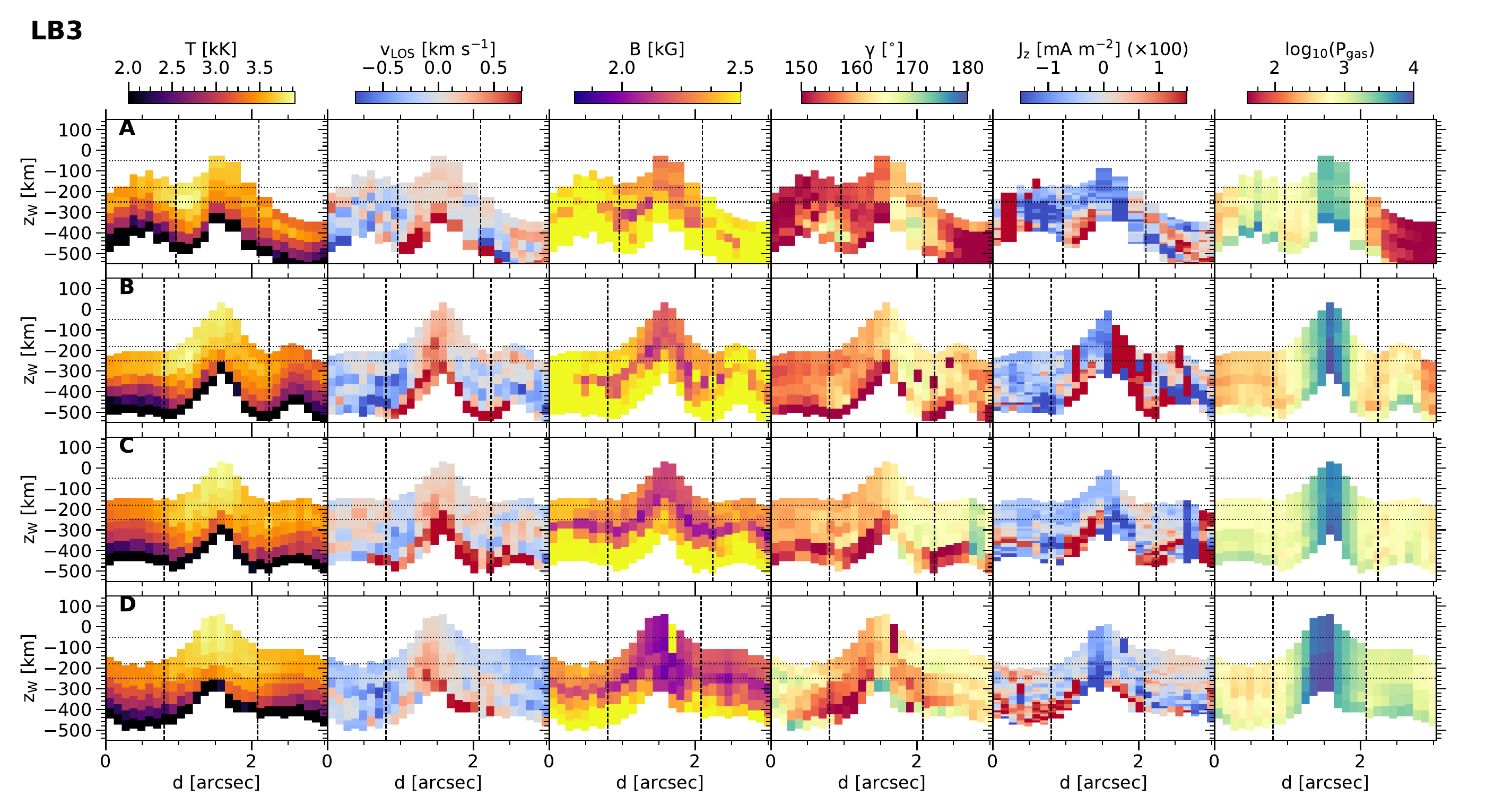}
\caption{Same as Fig.~\ref{fig:param_vertcut_lb1}, but for LB3. The horizontal lines indicate the heights shown in the bottom panels of Fig.~\ref{fig:param_z_lbs_mono}.}
\label{fig:param_vertcut_lb3}
\end{figure*}

Therefore, the height stratification along LB2 relies on whether a hot patch is present or not. Compared to positions outside them, hot patches host larger gradients of the inferred parameters along a seemingly narrower height range. According to the sensitivity of the observed \ion{Fe}{i} lines to the optical depth, the floor of LB2 at the hot patches is 300~km above that of the nearby umbra (found at around $-$350~km), and 150~km less deep than in the rest of the LB. The maximum heights reached along LB2 also differ, with the locations outside hot patches being the highest. 

\subsection{Light bridge 3}
\label{sub:complete_lb3}

Figure~\ref{fig:param_vertcut_lb3} shows the height stratification inferred for LB3 and confirms that this structure is elevated compared to its surroundings. Although it is very faint, we found consistent patterns across LB3, which reinforces the idea that the inferred results are reliable. 

The first thing to note about LB3 is that it shows differences on both sides. Specifically, the left side shows higher temperatures and more redshifted LOS velocities, having a difference of about 400~K and 0.2~km~s$^{-1}$ with respect to the adjacent umbra. Although the field inclination suggests rather vertical fields, the left side usually hosts more inclined fields, as mentioned in Fig.~\ref{fig:param_z_lbs_mono}. At the same time, compared to the surroundings, the magnetic fields at the center of LB3 are 300~G weaker, while the logarithm of the gas pressure is about 0.8 higher. We find mainly negative values for $J_z$, except in cut B, which shows a patch of strong positive values. Given that the magnetic field is very vertical, and the vertical component of the electric currents uses the horizontal components of the magnetic field for its calculation, we estimate that the inferred variations may not be reliable. 

In general, as height increases the temperature of LB3 rises, while the LOS velocity, magnetic field strength, and gas pressure decrease and the field inclination barely changes. These trends differ from the height gradients found in LB1 and LB2. The inferred signatures in LB3 do not seem to indicate the presence of convective flows, though results for the gas pressure may be compatible with a material supply from below. In addition, LB3 harbors the strongest magnetic fields (of about 2~kG) found in the analyzed LBs. Therefore, we suspect a possible mixing of information between LB3 and the surrounding umbra, which may hamper a better characterization of this LB.

\begin{figure}[!t]
\centering
\includegraphics[width=0.48\textwidth, trim={0.3cm 1.7cm 0.0cm 0.9cm}, clip]{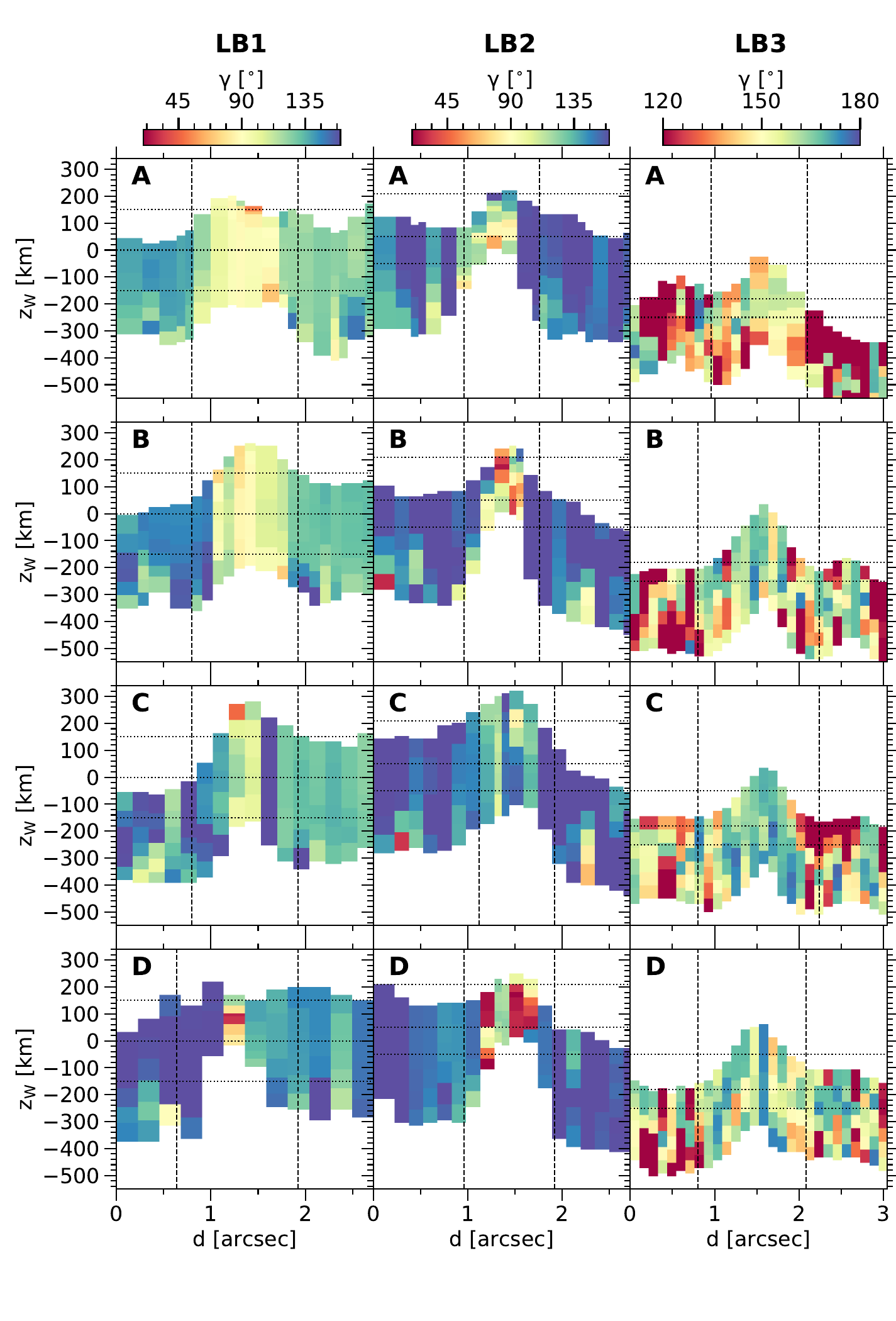}
\caption{Variation in the magnetic inclination given by SIR with geometric height along the cuts A--D in LB1, LB2, and LB3. The horizontal lines mark the heights displayed in Fig.~\ref{fig:param_z_lbs_mono} for LB1, LB2, and LB3. The vertical lines help distinguish the pixels within each LB. We note that the right column is saturated using a different scale.}
\label{fig:inc}
\end{figure}

\section{The height variation of the field inclination} \label{sub:inc}

By applying the SICON code, we obtained valuable information to analyze the height variation of diverse parameters in LBs. However, further discussion about the field inclination is missing. The SICON code infers the transverse and longitudinal magnetic fields as two independent parameters, so small errors in one of these components can lead to large errors in the inclination angle of the magnetic field vector. We are currently working on a version where the inclination is also involved in the training process, which would mitigate this problem.

In Sect.~\ref{sec:comp_sir} we describe how the inclination maps given by SIR show the LBs more conspicuously than those obtained with SICON. Thus, in this subsection we analyze the field inclination of the studied LBs using the field inclination and the Wilson depression given by SIR and SICON, respectively. We converted the field inclination values from the optical depth scale to geometric height following the same method that we applied to the SICON outputs (see the beginning of Sect.~\ref{sec:resgeo}).

Figure~\ref{fig:inc} shows the height variation of the field inclination along the cuts A--D in LB1, LB2, and LB3. For LB1 we observe a similar pattern to that obtained with SICON (see Fig.~\ref{fig:param_vertcut_lb1}), though the center of the LB shows more horizontal fields when using SIR. Furthermore, we only detect slight reversals in some positions (at the top of cuts A--C and in cut D). 

In LB2 the center tends again to show more horizontal fields, not only inside hot patches (cuts A, B, and D), but also outside them (cut C). In addition, only cut A shows a variation to more vertical fields with increasing height, which may indicate the presence of a cusp-like canopy. We also detect abrupt reversals in cuts B and D. 

Finally, LB3 shows more significant differences between the field inclination given by SICON and SIR, though we do not find a clear height variation in either case. Using SIR, we obtain rather vertical fields instead of more or less vertical fields on either side of the LB (see Fig.~\ref{fig:param_vertcut_lb3}).

The field inclination retrieved with SIR usually provides more accurate physical information in positions where we infer imprecise values with SICON. Furthermore, some positions in the LBs show three-lobed Stokes $V$ profiles formed by two antisymmetric $V$ profiles with opposite polarities and different lineshifts (examining deconvolved data). These $V$ profiles are related to the coexistence of magnetic components with opposite polarities along the LOS. Specifically, we find changes of polarity at positions of the lateral edges of LB2 with such profiles (cuts B and D in Fig.~\ref{fig:inc}), which may denote the presence of field reversals \citepads[as in][]{2016A&A...596A..59F}. However, other polarity reversals in the LBs appear at positions having more complex Stokes $V$ profiles. 

The detection of irregular Stokes $V$ profiles in LBs indicates significant changes in the physical properties between neighboring pixels \citepads[e.g.,][]{2014A&A...567A..96L, 2015A&A...584A...1L}. In particular, the lateral edges of LBs are squeezed between two environments of radically different physical properties, so their surroundings strongly affect the information from such positions. We plan to perform a future study using data simultaneously acquired in different photospheric lines at higher spatial resolution to determine the physical scenario in LBs, paying particular attention to their lateral edges.

\section{Discussion} \label{sec:discussion}

One of the challenges in our study lies in attempting to define a suitable model to explain the features inferred in the LBs. While LBs share certain generic properties, some essential features differ, as other investigations have also pointed out. It is not straightforward to define such a model since important aspects must be considered. First, the analyzed LBs were at different stages of their evolution and had different morphologies. In addition, one should take into account that using data acquired with a slit instrument imposes strong limitations when analyzing dynamic structures, such as LBs \citepads[e.g.,][]{2002A&A...383..275H, 2003ApJ...589L.117B, 2008SoPh..252...43L}. 

Light bridges come across as mountain ridges standing out from the umbra, which corroborates previous studies. The maximum height reached by each LB differs, with the umbral LB (LB3) being the deepest. In general, the analyzed LBs cover height ranges of 200--500~km, similar to the estimate given by \citetads{2004SoPh..221...65L}. Considering the sensitivity of the observed spectral lines in optical depth, the thickness of the height stratification seems to vary along the length of the LBs. Locations with hot patches, such as the southern part of LB1 and LB2, usually have narrower height stratifications compared to other positions. On the other hand, the thickness of LB3 barely changes along the structure, likely because of a mixing of information from LB3 and the umbra.  

Light bridges show clear variations in physical quantities with geometric height. Sometimes, we observe differences in the spatial distribution of some parameters when considering either optical depth or geometric height scales. This fact is of great interest, for instance, in the analysis of LB2, where using a geometric height scale has led to the characterization of its grainy nature.

Commonly, LBs are identified as bright protrusions that host convective motions of weakly-magnetized plasma compared to the nearby umbra. Our results show local signatures in each LB suggesting an injection of hot plasma into the LBs from below. These local signatures are usually the combination of a LOS velocity pattern compatible with the presence of convective motions, higher gas pressure, and weaker magnetic fields at specific positions. Thus, the injection of plasma may occur at these favored locations.  

Specifically, the hot patches host strongly reduced magnetic fields as a general rule, while the magnetic field in the northern part of LB1 is still very strong. Despite some differences, the northern part of LB1 shows patterns that are similar to those in penumbral filaments in the center-side penumbra. For instance, it shows a flow channel with a central blueshift bordered by redshifts along its length and a strong redshift at the tail. However, this flow channel penetrates from the limb-side penumbra to the umbra, which may suggest that this part of the LB is due to plasma protruding from the penumbra to the umbra through a conduit where the magnetic field is weaker than in the umbra. Under these conditions, an inflow to the umbra produced mainly by the gas pressure gradients along the magnetic field would show a LOS velocity of $-$1 -- $-$1.8~km~s$\mathrm{^{-1}}$ at the center of the structure, which is consistent with the observed values.
 
\begin{figure}[!t]
\centering
\includegraphics[width=0.48\textwidth, trim={1cm 2.5cm 1cm 2.5cm}, clip]{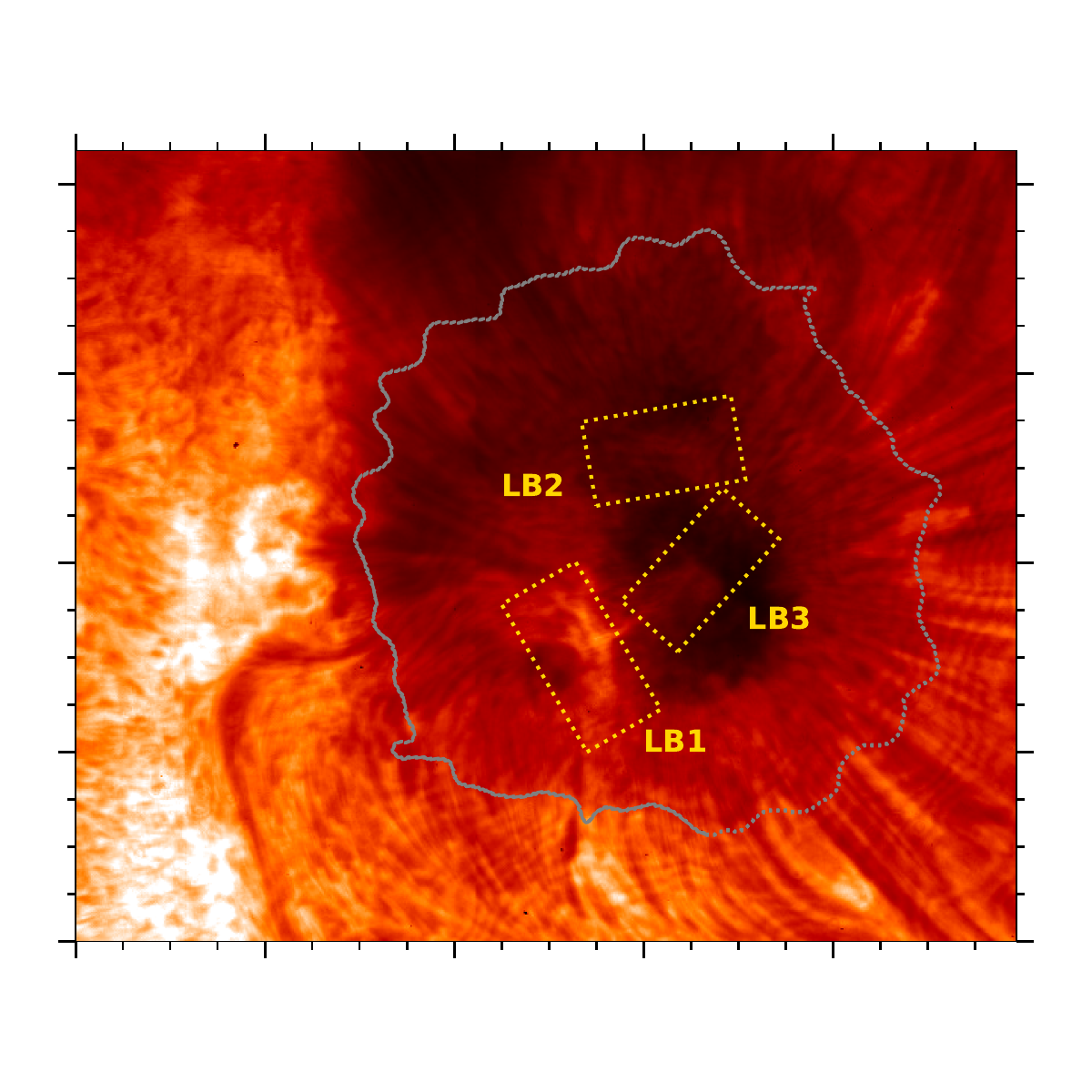}
\caption{Context H-$\mathrm{\alpha}$ image of the sunspot hosting the analyzed LBs at 19:06UT. The yellow rectangles enclose each LB. The gray contour delimits the outer boundary of the sunspot, as seen in the \ion{Fe}{i}~630.15~nm continuum intensity map. Each major tickmark represents 20\arcsec.}
\label{fig:contexto_lb1}
\end{figure}

Interestingly, the penumbral region near the northern part of LB1 seems connected to a filament existing in the above chromosphere, which has been studied by, for example, \citetads{2009ApJ...697..913O} and \citetads{2016A&A...589A..31B}. Figure~\ref{fig:contexto_lb1} shows this filament as seen in a H-$\mathrm{\alpha}$ image obtained with Hinode/SOT at the time the slit was scanning LB1. This filament may disturb the penumbra, which has a similar curvature to that of the LB and shows conspicuous signals in far-wing magnetograms \citepads[see Fig.~1 in][]{2010ASSP...19..193B}. Therefore, the inflow of plasma to the umbra in the northern part of LB1 could be related to the presence of this filament, as also mentioned by \citetads{2011ApJ...738...83S}. Similar behaviors in structures protruding into the umbra and linked to chromosphere activity have been inspected in other studies \citepads[e.g.,][]{2013ApJ...770...74K, 2017ApJ...846L..16G, 2019ApJ...880...34G}.

Diverse aspects of LB1 have been thoroughly analyzed in previous investigations, for example, in \citetads{2008SoPh..252...43L}, \citetads{2009ASPC..415..148S}, and \citetads{2011ApJ...738...83S}. Our results are generally compatible with those obtained in these studies, but some differences exist. Using data acquired almost one day later, \citetads{2009ApJ...704L..29L} found supersonic downflows at some locations of the northern part of LB1 that could be related to reconnection events. We have not inferred supersonic downflows, possibly because they were not present at the time the slit was scanning or because of differences in the inversion strategy. Nevertheless, Fig.~\ref{fig:contexto_lb1} shows a jet-like structure in the chromosphere above the northern part of LB1, specifically overlying the lateral redshifts found on its right side (see cut B in Fig.~\ref{fig:param_vertcut_lb1}). This coincidence may suggest a relation between the two imprints, which were possibly caused by reconnection events \citepads[as in][]{2009ApJ...704L..29L}. Furthermore, we also observe significant divergences in the field inclination values we obtained in the northern part of LB1 and those reported by \citetads{2009ASPC..415..148S} and \citetads{2011ApJ...738...83S}. We retrieve mainly horizontal fields using the SIR code (or even of opposite polarity to that of the sunspot with SICON) flanked by more vertical fields of 120--130$\mathrm{^\circ}$ inclination, while these authors found average inclination values of 130--140$\mathrm{^\circ}$ using Milne-Eddington inversions. Such a difference is probably due to the strategy employed during the inversions. In contrast, our results in the field inclination are consistent with those obtained by \citetads{2009ApJ...704L..29L} with SIR.

The analyzed LBs also harbor enhanced electric current densities, specifically at positions with strong gradients in the magnetic field strength and inclination. However, the abrupt changes found in this latter parameter in the southern part of LB1 and LB2 may affect the reliability of the current estimates there. The horizontal component of the electric current density vector dominates over the vertical component, as also found by \citetads{2021A&A...652L...4L} in an LB and by \citetads{2010ApJ...721L..58P} in the penumbra. However, it is $J_z$ the component that differs significantly among the LBs, with the northern part of LB1 showing the greatest values. In general, both the enhanced $J_h$ and $J_z$ components partially overlap regions with higher temperatures, so they appear as a shell that is elongated or patchy depending on the shape of the hotter regions in the LBs.

According to \citetads{2011ApJ...738...83S}, the positive and negative enhanced $J_z$ values in the northern part of LB1 may have different natures depending on their position in the LB. On April 30 (about ten hours before observing the data we used), \citetads{2009ASPC..415..148S} found positive $J_z$ values almost at the center of the LB and negative ones along a lateral edge. They interpreted the former as field-aligned currents along the magnetic flux tube lying in the LB, but the latter were thought to be currents induced by the sudden change in the field orientation between the LB and the nearby umbra. However, \citetads{2011ApJ...738...83S} used this latter interpretation to explain both the positive and negative strong currents on May 1, as they appeared on either side of the LB. Our study thus gives an intermediate picture between those described by \citetads{2009ASPC..415..148S} and \citetads{2011ApJ...738...83S}.

Positions with strong $J_z$ in the northern part of LB1 seem related to chromospheric brightenings detected above this region (see Fig.~\ref{fig:contexto_lb1}). Such brightenings manifest a local temperature increase in the overlying chromosphere that may be due to current heating. As a first approach, we have estimated that such enhanced currents lead to a Joule heating of about 10$\mathrm{^{-1}}$~erg~s$\mathrm{^{-1}}$~cm$\mathrm{^{-3}}$. In particular, the vertical currents are related to a Joule heating that is ten times lower, and one order of magnitude greater than that given by \citetads{2006A&A...453.1079J}. We used the expressions $|J|^\mathrm{2}/\sigma$ and $J_z^\mathrm{2}/\sigma$ to compute each Joule heating value, with $\sigma$ being the electric conductivity estimated following \citetads{1969SoPh....6..241K}. In the near future, we plan to investigate the temporal evolution of LB1 and its relation with the nearby filament and the overlying chromospheric activity by applying the SICON code. As demonstrated here, this code offers an easy way to estimate the height variation of diverse parameters and, specifically, of the electric current density vector, which is needed to determine the contribution of the ohmic heating to the chromospheric brightenings.

For a general view, we also compare our results with studies related to other LBs. The different stratifications inferred by us are consistent with the diverse behaviors previously reported. For instance, the northern part of LB1 shows a horizontal (or slightly reverse) magnetic field that weakens with increasing height. This height variation is compatible with the findings of \citetads{2021A&A...647A.148G}. On the other hand, LB2 and the southern part of LB1 show an increase in the magnetic field strength with height, while the temperature becomes lower in agreement with  \citetads{2006A&A...453.1079J}, \citetads{2016A&A...596A..59F}, and \citetads{2021A&A...647A.190B}, among others. Furthermore, we corroborate that magnetic fields in LBs are usually more horizontal than in the nearby umbra and, in some cases, they become more vertical at the lateral edges of the LB. In this regard, while some studies proposed cusp-like magnetic canopies over LBs, we found only two positions that support such a scenario. However, we cannot rule out the existence of a magnetic canopy possibly due to insufficient spatial resolution. Finally, some positions in LBs reveal field reversals that may be real (as in \citealt{2014A&A...568A..60L} and \citealt{2016A&A...596A..59F}) since they coincide with deconvolved Stokes $V$ profiles of three lobes, which suggest the presence of fields with opposite polarity along the LOS. However, we note that the information from these particular locations may be affected by that from nearby pixels due to the deconvolution process.

\section{Summary and conclusions}

In this work we studied the physical scenario of three LBs in terms of the geometric height. To this end, we used the capability of the SICON code to provide, along with the thermodynamic and magnetic properties, the Wilson depression at each optical depth and pixel.

Filamentary and grainy LBs reveal conspicuous differences when inspecting their height stratification. Considering the sensitivity of the observed \ion{Fe}{i} lines, the former suggests the presence of a horizontal plasma inflow to the umbra, which may be due to a filament anchored nearby. On the other hand, regions with grainy morphologies reveal specific positions compatible with convective flows of hot plasma injected from below. The latter, moreover, shows larger gradients in the physical parameters along narrower height ranges. 

The height variation of the magnetic field is usually incompatible with the magnetic canopy scenario. The reason could be a lack of spatial resolution or the inversion strategy adopted here. However, this result might indicate that a magnetic canopy exists only at specific positions, which could agree with the detection of LB-like structures in the chromosphere and TR \citepads[][]{2018A&A...609A..73R}. Furthermore, shells of enhanced electric current density values partially overlap with positions with strong variations in the magnetic field strength and inclination, giving the impression that they envelop the LBs. Our estimates corroborate previous results in general. However, we also find differences in each LB, not only regarding their distribution, but also their values. In this regard, we infer the strongest $J_z$ values in the filamentary LB, where the positive currents are particularly significant and seem related to brightenings and a jet-like ejection in the above chromosphere.

An appealing facet of our study is that it provides a springboard to conduct future investigations. For instance, we plan to investigate how the properties of LB1 are related to the chromospheric events detected above. Furthermore, multiwavelength spectropolarimetric data taken at high spatial resolution with the next-generation large-aperture telescopes, such as DKIST \citepads{2020SoPh..295..172R} and EST \citepads{2022A&A...666A..21Q}, will offer a golden opportunity for analyzing the most evasive aspects of LBs.

Finally, we emphasize the capability of SICON to provide results consistent with those previously reported in LBs. Since it provides a straightforward manner to infer the physical parameters in terms of the geometric height, the application of this code is of great interest to analyze the atmospheric stratification in other solar structures.

\begin{acknowledgements}
We acknowledge the funding received from the European Research Council (ERC) under the European Union's Horizon 2020 research and innovation program (ERC Advanced Grant agreement No. 742265) and from the Agencia Estatal de Investigación del Ministerio de Ciencia, Innovación y Universidades (MCIU/AEI) under grant ``Polarimetric Inference of Magnetic Fields'' and the European
Regional Development Fund (ERDF) with reference
PID2022-136563NB-I00/10.13039/501100011033. AAR acknowledges financial support from the Spanish Ministerio de Ciencia, Innovaci\'on y Universidades through project PGC2018-102108-B-I00 and FEDER funds. This research is supported by the Research Council of Norway, project number 325491 and through its Centers of Excellence scheme, project number 262622. This work has been partially funded by AEI/MCIN/10.13039/ 501100011033/PID2021-125325OB-C55 and European REDEF funds ("A way of making Europe"). Hinode is a Japanese mission developed and launched by ISAS/JAXA, collaborating with NAOJ as a domestic partner, NASA and STFC (UK) as international partners. Scientific operation of the Hinode mission is conducted by the Hinode science team organized at ISAS/JAXA. This team mainly consists of scientists from institutes in the partner countries. Support for the post-launch operation is provided by JAXA and NAOJ (Japan), STFC (UK), NASA, ESA, and NSC (Norway). This research has made use of NASA's Astrophysical Data System. We thank J. de la Cruz Rodríguez for his GitHub repository, we used the \texttt{witt} package in this work. We acknowledge the community effort devoted to the development of the following open-source packages that we used in this work: \texttt{numpy} \citepads{harris2020array} and \texttt{matplotlib} \citep{hunter:2007}. This work made use of \texttt{Astropy}: a community-developed core Python package and an ecosystem of tools and resources for astronomy \citepads{2022ApJ...935..167A}. We would also like to thank the anonymous referee for helpful comments.
\end{acknowledgements}


\end{document}